\documentstyle[12pt]{article}

\newcommand{\rf}[1]{(\ref{#1})}

\def\be{\begin{equation}}
\def\ee{\end{equation}}

\begin{document}

%\draft

\begin{flushright}
hep-th/0201226
\\
FIAN/TD/02-01
\end{flushright}

\vspace{1cm}

\begin{center}

{\Large\bf  Massless arbitrary spin fields in $AdS_5$}

\vspace{2.5cm} R.R. Metsaev

\vspace{1cm} {\it Department of Theoretical Physics, P.N. Lebedev
Physical Institute,

\medskip
Leninsky prospect 53, 119991, Moscow, Russia}

\vspace{3.5cm} {\bf Abstract}

\end{center}

{}Free arbitrary spin massless and self-dual massive fields
propagating in $AdS_5$ are investigated. We study totally
symmetric and mixed symmetry fields on an equal footing.
Light-cone gauge action for such fields is constructed. As an
example of application of light-cone formalism we discuss
$AdS/CFT$ correspondence for massless arbitrary spin $AdS$ fields
and corresponding boundary operators at the level of two point
function.

\newpage

\renewcommand{\thefootnote}{\arabic{footnote}}
\setcounter{footnote}{0}

{\bf Introduction}. A study of higher spin gauge theory has two
main motivations (see e.g. \cite{vas1,met8}): Firstly to overcome
the well-known barrier of $N\leq 8$ in $d=4$ supergravity models
and, secondly, to investigate if there is a most symmetric phase
of superstring theory that leads to the usual string theory as a
result of a certain spontaneous breakdown of higher spin
symmetries. Another motivation came recently from conjectured
duality of free large $N$ conformal ${\cal N}=4$ SYM theory and a
theory of massless higher spin fields in $AdS_5$
\cite{hagsun,sun}. Discussion of this theme in the context of
various limits in $AdS$ superstring may be found in
\cite{ts1,sez1}. In both the tensionless superstring theory and
massless higher spin fields theory all types of massless fields
appear in general. This implies that all types of massless fields
should be studied. This is what we are doing in this paper.

Let us first formulate the main problem we solve in this letter.
Fields propagating in $AdS_5$ space are associated with
positive-energy unitary lowest weight representations of $SO(4,2)$
group. A positive-energy lowest weight irreducible representation
of the $SO(4,2)$ group denoted as $D(E_0,{\bf h})$, is defined by
$E_0$, the lowest eigenvalue of the energy operator, and by ${\bf
h}=(h_1,h_2)$, $h_1 \geq |h_2|$, which is the highest weight of
the unitary representation of the $SO(4)$ group. The highest
weights $h_1,h_2$ are integers and half-integers for bosonic and
fermionic fields respectively. In this paper we investigate the
fields whose $E_0$ and ${\bf h}$ are given by

\begin{eqnarray}\label{e01} && E_0 = h_1 + 2\,, \qquad  h_1 > |h_2|\,,
\\
\label{e02} && E_0> h_1 + 1\,,  \ \ \ \ \ h_1 = |h_2|\,,\qquad
|h_2|
 > 1/2\,. \end{eqnarray} The fields in \rf{e01} are massless
fields while the ones in \rf{e02} are massive self-dual fields.
The massless fields with $h_1\geq 1$, $h_2=0,\pm 1/2$ are referred
to as totally symmetric fields while both the massless and massive
fields with $|h_2|> 1/2$ we refer to as mixed symmetry fields. In
manifestly Lorentz covariant formulation the totally symmetric
and mixed symmetric fields are the tensor fields whose $SO(4,1)$
space-time indices have the structure of the respective Young
tableauxes with one and two rows\footnote{We note that the case
${\bf h}=(1,0)$ corresponds to spin one Maxwell field,  the case
${\bf h}=(2,0)$ is the graviton, and the cases ${\bf h}=(1,\pm1)$
correspond to 2-form potentials.}. Actions for the massless
totally symmetric integer and half-integer spin fields in $AdS_d$
space were found in \cite{lopvas,vasfer}. The gauge invariant
equations of motion for all types of massless fields (totally and
mixed symmetric ones) in $AdS_d$ space for even $d$ have been
found in \cite{metsit1,mettal}\footnote{ The particular case of
massless field corresponding to ${\bf h}=(2,1)$, for $d\geq 6$
was described at the action level in \cite{brimet}. Gauge
invariant actions for $AdS_4$ massless fields have been
established in \cite{f1,ff}. Various gauge invariant descriptions
of totally symmetric massless fields in terms of the potentials
in $AdS_d$ may be found in \cite{rrm1,seg,buc}. Massive self-dual
arbitrary spin fields in $AdS_3$ were investigated in
\cite{rrm4}. Discussion of massive totally symmetric fields in
$AdS_d$, $d\geq 4$, may be found in \cite{des,zin}.}. Massless
mixed symmetry $AdS_5$ fields with arbitrary ${\bf h}$ and
self-dual massive fields with arbitrary $E_0$ have not been
described at the field theoretical level so far \footnote{ Group
theoretical description of representation of $so(4,2)$ algebra
for arbitrary ${\bf h}$ and discrete values of $E_0$ via
oscillator method \cite{bargun} may be found, e.g., in
\cite{gunmin}. Lorentz covariant equations of motion for $AdS_5$
self-dual massive fields with special values of $E_0$ were
discussed in \cite{sez2}.}. In this paper we develop a light-cone
gauge formulation for such fields at the action level. Using the
light-cone gauge formalism in $AdS$ space developed in
\cite{rrm1}, we describe both the totally and mixed symmetry
fields on an equal footing. Since, by analogy with flat space, we
expect that a quantization of the Green-Schwarz $AdS$ superstring
with Ramond-Ramond charge will be available only in the light-cone
gauge \cite{mt3,mtt} it seems that from the stringy perspective of
$AdS/CFT$ correspondence the light-cone approach is the fruitful
direction to go.

\medskip
{\bf Light-cone gauge action}. Let $\phi(x)$ be a arbitrary spin
field propagating in $AdS_5$ space. If we collect spin degrees of
freedom in a generating function $|\phi(x)\rangle$ then a
light-cone gauge action for this field can be cast into the
following `covariant form'\cite{rrm1}\footnote{ We use
parametrization of $AdS_5$ space in which
$ds^2=(-dx_0^2+dx_i^2+dx_3^2+dz^2)/z^2$. Light-cone coordinates
in $\pm$ directions are defined as $x^\pm=(x^3 \pm x^0)/\sqrt{2}$
and $x^+$ is taken to be a light-cone time. Unless otherwise
specified, we adopt the conventions: $i,j =1,2$.
$\partial_i\equiv\partial/\partial x^i$,
$\partial_z\equiv\partial/\partial z$, $\partial^\pm=\partial_\mp
\equiv \partial/\partial x^\mp$. We use indices $I,J$ for $so(3)$
tangent space vectors $X^I=(X^i,X^z)$. Sometimes, instead of
$so(2)$ vector indices $i,j=1,2$ we use complex frame indices
$i,j=x,\bar{x}$. In this notation a $so(2)$ vector $X^i$ is
representable by $X^x$, $X^{\bar{x}}$ and for scalar product we
have the decomposition $X^iY^i= X^xY^{\bar{x}}+X^{\bar{x}}Y^x$.}

\be\label{2lcact} S_{l.c.} =\frac{1}{2}\int d^5x \langle
\phi(x)|\bigl(\Box -\frac{1}{z^2}A\bigr)|\phi(x)\rangle\,, \qquad
\Box = 2\partial^+\partial^- + \partial_i^2+\partial_z^2\,, \ee
where an $A$ is some operator not depending on space-time
coordinates and their derivatives. This operator acts only on
spin indices of fields collected in ket-vector $|\phi\rangle$.
We   call the operator $A$ the $AdS$ mass operator. In the
remainder of this paper we find it convenient to Fourier
transform to momentum space for all coordinates except for the
radial $z$ and the light-cone time $x^+$:

\be |\phi(x^+,x^-,x^i,z)\rangle =\int
\frac{d^{2}pd\beta}{(2\pi)^{3/2}} e^{{\rm i}(\beta x^- + p^i
x^i)} |\phi(x^+,\beta,p^i,z)\rangle\,, \ee where we use $\beta$ to
denote a momentum in light-cone direction: $\beta\equiv p^+$. Now
we rewrite the action in the Hamiltonian form

\be\label{lcaction} S_{l.c.} = \int dx^+ d^2p\,d\beta\,dz
\langle\phi(x^+,-p,z)| \beta ({\rm i}\partial^- +
P^-)|\phi(x^+,p,z)\rangle\,, \ee where $P^-$ is the (minus)
Hamiltonian density

\be\label{pm1} P^-=\frac{-p_i^2 + \partial_z^2}{2\beta} -
\frac{1}{2z^2\beta}A\,. \ee In Eq. \rf{lcaction} and below a
momentum $p$ as argument of wave function $\phi(x^+,p,z)$
designates the set $\{p^i,\beta\}$. All that remains to fix
action explicitly is to find the $AdS$ mass operator. The
explicit form of this operator depends on what kind of
realization of spin degrees of freedom we use. Let us first
consider the bosonic fields.

\medskip
{\bf Bosonic fields}. To describe light-cone gauge formulation of
the $D(E_0,{\bf h})$ massless or self-dual massive bosonic field
$|\phi\rangle$ we introduce a {\it complex-valued} totally
symmetric traceless $so(3)$ tensor field $\phi^{I_1\ldots
I_{h_1}}$. As usual to avoid cumbersome tensor expressions we
introduce creation and annihilation oscillators $\alpha^I$ and
$\bar{\alpha}^I$

\be\label{bososc} [\bar{\alpha}^I,\,\alpha^J]=\delta^{IJ}\,,\qquad
\bar{\alpha}^I|0\rangle =0\,, \ee and the generating function
$|\phi\rangle$ defined by \be\label{genfun1} |\phi(x^+,p,
z)\rangle \equiv \alpha^{I_1}\ldots \alpha^{I_{h_1}}
\phi^{I_1\ldots I_{h_1}}(x^+,p\,, z)|0\rangle\,. \ee

The $AdS$ mass operator is found then to be

\be\label{adsmas1} A =-\frac{1}{2}m^{ij}m^{ij} -
\frac{1}{4}\,,\qquad m^{ij}\equiv M^{ij} -{\rm
i}\epsilon^{ij}c\,,\ee $\epsilon^{12} = - \epsilon^{21}= 1$,
where the $so(2)$ spin operator $M^{ij}$ takes the standard  form
\be\label{spiope0} M^{ij} = \alpha^i\bar{\alpha}^j
-\alpha^j\bar{\alpha}^i\,,\ee and the number $c$  is given by

\be\label{c12} c = \left\{\begin{array}{cl} h_2\,, &
\hbox{ for massless fields \rf{e01}}\,, \\[8pt]
(E_0 - 2)\hbox{sign }h_2\,, &\hbox{ for massive self-dual fields
\rf{e02}}\,,
\end{array}\right.
\ee where $\hbox{sign} h = +1(-1)$ for $h>0$($h<0$). Details of
derivation may be found in Appendix B. A few comments are in
order.

i) The light-cone gauge action for the mixed symmetry $D(E_0,{\bf
h})$ field, ($|h_2| >0$), is formulated in terms of the {\it
complex-valued} rank $h_1$ totally symmetric traceless $so(3)$
tensor field. Number of physical degrees of freedom ($DoF$) of
such field is equal to $2(2h_1+1)$, i.e., for the field  in
\rf{e01}, it depends only on $h_1$ and does not on $h_2$.

(ii) {}For the case of massless totally symmetric field ($h_2=0$)
we can impose reality condition on the complex-valued $so(3)$
tensor field and this gives the standard light-cone description
of the massless totally symmetric field \cite{rrm1}. Number of
physical $DoF$ of such field is equal to $2h_1+1$.

(iii) The $h_2$ of $AdS_5$ fields is similar to helicity of
massless fields  in flat four dimensional space. As is well known
number of physical $DoF$ of the latter fields does not depend on
helicity and is equal to two. If $AdS_5$ field $|\phi\rangle$ has
helicity $h_2$ then its complex-conjugated
  $\langle \phi |$  has opposite helicity $-h_2$, i.e.,
the light-cone gauge action \rf{lcaction} is formulated in terms
of fields having opposite values of $h_2$.

\medskip {\bf Fermionic fields}. In this case $h_1$ and $h_2$
are half-integers and in what follows we use notation $h_1^\prime
=h_1 -\frac{1}{2}$. To discuss light-cone gauge formulation of
$D(E_0,{\bf h})$ fermionic field we introduce complex-valued
traceless $so(3)$ spin-tensor field $\phi_\gamma\!\!{}^{I_1\ldots
I_{h_1' }}$, $\gamma =1,2$, which is symmetric in $I_1,\ldots,
I_{h_1'}$. In addition to the bosonic creation and annihilation
oscillators \rf{bososc} we introduce fermionic oscillators
$\eta^\gamma$, $\bar{\eta}^\gamma$

\be\label{ferosc2}
\{\bar{\eta}^\alpha,\eta^\gamma\}=\delta^{\alpha\gamma}\,,\qquad
\bar{\eta}^\gamma |0\rangle =0\,.\ee For this case the Fock
vector $|\phi\rangle$ defined by

\be\label{genfun2} |\phi(x^+,p, z)\rangle\equiv \alpha^{I_1}\ldots
\alpha{}^{I_{h_1^\prime}}\eta^\gamma \phi_\gamma{}\!\!^{I_1\ldots
I_{h'_1 } }(x^+,p , z)|0\rangle \ee is assumed to satisfy the
constraint

\be (\bar{\alpha}^i \sigma^i
-\bar{\alpha}^z)\sigma^3\bar{\eta}|\phi\rangle =0\,,
  \ee where $\sigma^i$, $\sigma^3$ are
Pauli matrices. This constraint amounts to requiring the
spin-tensor field  $ \phi_\alpha{}\!\!^{I_1\ldots I_{h'_1 } }$
transforms in irreducible spin $h_1$ representation of $so(3)$
algebra. The operator $A$ and the number $c$ take then the form
given by formulas \rf{adsmas1},\rf{c12} in which the spin operator
$M^{ij}$ \rf{spiope0} should be replaced by

\be\label{spiope1n} M^{ij} = \alpha^i\bar{\alpha}^j
-\alpha^j\bar{\alpha}^i + \frac{\rm
i}{2}\epsilon^{ij}\eta\sigma^3\bar{\eta}\,. \ee

As a side of remark we note that these results can be presented
in an equivalent form which is somewhat convenient for evaluation.
Instead of tensor fields \rf{genfun1},\rf{genfun2} we can
introduce $2h_1+1$ complex-valued fields $\phi_m$ which with the
help of a complex-valued variable $\zeta$ we collect in the
$|\phi\rangle$ as

\be\label{zetexp} |\phi(x^+,p,z)\rangle = \sum_{m = -h_1}^{h_1}
\zeta^{h_1 - m}\phi_m(x^+,p,z)\,,   \ee where for bosons
$m=0,\pm1,\ldots ,\pm h_1$, while for fermions $m=\pm
1/2,\pm3/2,\ldots ,\pm h_1$. The $\zeta$, which we refer to as
projective variable, is used to formulate results entirely in
terms of the generating function $|\phi\rangle$. The scalar
product for such generating functions is defined to be

\be\label{scapro} \langle \phi |\varphi \rangle =\int \frac{d\zeta
d\bar{\zeta}}{(1+\frac{1}{2}\zeta\bar{\zeta})^{2h_1+2}}
\overline{\phi (\zeta)}\varphi (\zeta)\,.\ee The $AdS$ mass
operator for this realization takes the form

\be\label{adsmas2} A =(\zeta\partial_\zeta + a)^2- \frac{1}{4}\,,
\qquad a = c - h_1\,,\ee where the number $c$ is given by
\rf{c12}. In the complex notation the spin operator $M^{ij}$ is
representable by $M^{x\bar{x}}$ which takes the form

\be\label{spiope3}\label{spiope3n} M^{x\bar{x}} = -\zeta
\partial_\zeta + h_1\,,\qquad \partial_\zeta \equiv \partial/\partial
\zeta\,. \ee We note that the representation for the operators $A$
\rf{adsmas2} and $M^{x\bar{x}}$ \rf{spiope3} is valid for both
the integer and half-integer fields, i.e., while using the
projective variable $\zeta$ it is not necessary to introduce
anticommuting oscillators \rf{ferosc2} to discuss the
half-integer fields.

\medskip
{\bf Global symmetries of light-cone gauge action}. We turn now
to discussion of global $so(4,2)$ symmetries of the light-cone
gauge action. The choice of the light-cone gauge spoils the
manifest global symmetries,  and in order to demonstrate that
these global invariances are still present one needs to find the
Noether charges which generate them\footnote{These charges play  a
crucial role  in formulating interaction vertices in field
theory. Application of Noether charges in formulating superstring
field theories may be found in \cite{gsb},\cite{gs1}}. The field
theoretical representation for Noether charges in $AdS$ light-cone
gauge formalism takes the standard form \cite{rrm1}

\be \hat{G} = \int dzd^{2}pd\beta\, \beta\langle \phi(x^+,-p,z)|
G|\phi(x^+,p,z)\rangle\,, \ee where $G$ are differential operators
acting on physical field $|\phi\rangle$. Noether charges
(generators) can be split into two groups: kinematical and
dynamical generators (see Appendix A). Representation of the
kinematical generators is given by
\begin{eqnarray}
\label{3spi} && P^i=p^i\,, \qquad P^+=\beta\,,
\\
&& D = {\rm i}x^+ P^- -\partial_\beta \beta -\partial_{p^i}p^i
+z\partial_z + \frac{3}{2}\,,
\\
\label{3sjpm} && J^{+-} = {\rm i}x^+P^- + \partial_\beta \beta\,,
\\
\label{3sjpi} && J^{+i}= {\rm i} x^+ p^i +
\partial_{p^i}\beta\,,
\\
&& J^{ij}= p^i\partial_{p^j} - p^j\partial_{p^i}+M^{ij}\,,
\\
\label{3skp} && K^+ = \frac{1}{2}(2{\rm i} x^+ \partial_\beta -
\partial_{p^i}^2 +z^2)\beta +  {\rm i} x^+ D\,,
\\
&& K^i =\frac{1}{2}(2{\rm i} x^+ \partial_\beta -
\partial_{p^j}^2 +z^2)p^i -\partial_{p^i} D - M^{ij}\partial_{p^j}
-M^{zi} z + {\rm i} M^{i-}x^+\,,
\end{eqnarray}
where $\partial_{p^i} \equiv \frac{\partial}{\partial p^i}$,
$\partial_\beta \equiv \frac{\partial}{\partial \beta}$. The
dynamical generator $P^-$  is given by \rf{pm1} while the
remaining ones are
\begin{eqnarray}
\label{genjmi} && J^{-i} = -\partial_\beta p^i +\partial_{p^i} P^-
+M^{-i}\,,
\\
\label{genkp} && K^-= \frac{1}{2} (2{\rm i} x^+ \partial_\beta -
\partial_{p^i}^2 +z^2)P^- -\partial_\beta D
+\frac{1}{\beta}p^i\partial_{p^j} M^{ij} +\frac{1}{\beta}(zp^i
-\partial_z\partial_{p^i}) M^{zi}
\nonumber\\
&&\hspace{2cm} - \frac{\partial_{p^i}}{2z\beta}[M^{zi},A]
+\frac{1}{\beta}B\,,
\end{eqnarray}
where `nonlinear' spin operator $M^{-i}=-M^{i-}$ is given by \be
M^{-i} \equiv \frac{1}{\beta}\Bigl(M^{ij}p^j + M^{zi}\partial_z +
\frac{1}{2z}[M^{zi}, A]\Bigr)\,. \ee The operators $B$, $M^{zi}$
are acting only on spin degrees of freedom and are given by
\be\label{bope1} B = -M^{zi}M^{zi} +\frac{\rm
i}{2}c\epsilon^{ij}M^{ij}\,, \ee \be\label{spiope2n} M^{zi} =
\alpha^z\bar{\alpha}^i + \alpha^i\bar{\alpha}^z+
\frac{1}{2}\eta\sigma^i\bar{\eta}\,. \ee For the case of bosonic
fields we should simply ignore the fermionic term in expression
for $M^{zi}$ \rf{spiope2n}. We note that the spin operators
$M^{ij}$, $M^{zi}$ satisfy commutation relations of $so(3)$
algebra \be [M^{zi},M^{zj}]= M^{ij}\,, \qquad
[M^{ij},M^{zk}]=\delta^{jk}M^{zi}- \delta^{ik} M^{zj}\,, \ee and
a realization of the spin operator $M^{zi}$ in terms of
projective variable $\zeta$ takes the form \be\label{spiope4}
M^{zx}=\partial_\zeta\,,\qquad
M^{z\bar{x}}=-\frac{1}{2}\zeta^2\partial_\zeta + h_1\zeta\,. \ee
Since the light-cone gauge action \rf{lcaction} is invariant with
respect to the global symmetries generated by $so(4,2)$ algebra,
$\delta_{\hat{G}} |\phi\rangle = G|\phi\rangle$, the formalism we
discuss is  sometimes referred to as an off shell light-cone
formulation \cite{gsb}.

\bigskip

{\bf Ground state of $AdS_5$ fields}. In this section we would
like to demonstrate that the massless and self-dual massive fields
discussed above are indeed carriers of the positive-energy
unitary lowest weight representations of $SO(4,2)$ group labeled
by $D(E_0,{\bf h})$ where $E_0$ is given by Eqs.
\rf{e01},\rf{e02}. To do this we evaluate ground state of $AdS$
fields. All that requires is to demonstrate that $E_0$ for ground
states of massless and self-dual fields satisfy the basic
relations \rf{e01},\rf{e02}. To this end we begin with definition
of ground state $|v\rangle$ (vacuum). The most convenient way to
do this is to use $so(4,2)$ algebra taken to be in energy basis
(see Appendix A). In this basis one has energy operator
$J^{w\bar{w}}$, deboost operators $J^{wy}$, $J^{w\bar{y}}$,
$J^{wi}$, boost operators, $J^{\bar{w}y}$, $J^{\bar{w}\bar{y}}$,
$J^{\bar{w}i}$ and six operators, $J^{iy}$, $J^{i\bar{y}}$,
$J^{y\bar{y}}$, $J^{ij}$, which are generators of the $so(4)$
algebra\footnote{In complex notation $J^{ij}$ is representable by
$J^{x\bar{x}}$ while $so(2)$ vectors like $J^{wi}$ are
representable by $J^{wx}$ and $J^{w\bar{x}}$}. The vacuum forms a
linear space which is invariant under the action of the energy
operator $J^{w\bar{w}}$ and generators of the $so(4)$ algebra. In
other words the $|v\rangle$ is i) the eigenvalue vector of
$J^{w\bar{w}}$; ii) a weight ${\bf h}$ representation of the
$so(4)$ algebra. Now the vacuum $|v\rangle$ is defined by the
relations

\be\label{9} J^{w\bar{w}}|v\rangle= E_0|v\rangle\,, \ee
\be\label{10} J^{wy}|v\rangle=0\,,\qquad
J^{w\bar{y}}|v\rangle=0\,,\ee \be \label{11}
J^{wi}|v\rangle=0\,.\ee Eq. \rf{9} reflects the fact that the
$|v\rangle$ is the eigenvector of the energy operator
$J^{w\bar{w}}$, while Eqs. \rf{10},\rf{11} tell  us that $E_0$ is
the lowest value of the energy: since the deboost operators
decrease the energy they should evidently annihilate the
$|v\rangle$. Just as in the ordinary Poincar\'e quantum field
theory one can build the representation Fock space $D(E_0,{\bf
h})$ by acting with boost operators  $J^{\bar{w}y}$,
$J^{\bar{w}\bar{y}}$, $J^{\bar{w}i}$ on the vacuum states
$|v\rangle$.

Without loss of generality and to simplify our presentation we \
i) evaluate representative of ground state denoted by
$|v^{E_0,{\bf h}}\rangle$ which is a highest weight vector of
representation of the $so(4)$ algebra; ii)  evaluate ground state
for $x^+=0$ and $\beta>0$; iii) drop the $|v^{E_0,{\bf
h}}\rangle$ in equations and instead of equations like
$X|v^{E_0,{\bf h}}\rangle=Y|v^{E_0,{\bf h}}\rangle$ we write
simply $X\approx Y$. In addition to \rf{9}-\rf{11} the
$|v^{E_0,{\bf h}}\rangle$ satisfies, by definition, the following
constraints

\be \label{11n} J^{y\bar{y}} \approx h_1\,,\qquad \qquad
J^{x\bar{x}} \approx h_2\,,\ee \be  \label{12} J^{yi}\approx 0\,.
\ee Next step is to rewrite Eqs. \rf{9}-\rf{12} in terms of
generators in light-cone basis \rf{3spi}-\rf{genkp}. Making use of
\rf{11},\rf{12} and formulas \rf{in7},\rf{in9} we get the
equations \be\label{13}
 P^i+J^{+i}\approx 0\,,\ee
\be \label{14} K^i -J^{-i} \approx 0\,.\ee Using Eqs. \rf{10} and
formulas \rf{in3},\rf{in4} we can express $P^-$ and $K^-$ in
terms of $P^+$, $K^+$, $D$ and $J^{+-}$. Exploiting this
representation for $P^-$, $K^-$ in \rf{in1},\rf{in2} we get from
\rf{9} and the 1st equation in \rf{11n} the following equations

\be\label{eps} P^+ +\frac{1}{2}(-D +J^{+-}) \approx
\frac{1}{2}(E_0 + h_1)\,,\ee
 \be \label{ems} K^+ +\frac{1}{2}(D+J^{+-}) \approx \frac{1}{2}(E_0
- h_1)\,. \ee Now making use of concrete representation for
generators in terms of differential operators
\rf{3spi}-\rf{genkp} we can find ground state explicitly. Using
Eqs. \rf{13},\rf{eps},\rf{ems} we find respective dependence of
the ground state $|v^{E_0,\bf{h}}\rangle$ on momenta $p^i$,
$\beta$ and radial coordinate $z$:

\be\label{v10} |v^{E_0,\bf{h}}(p,\beta,z)\rangle =
\exp(-\frac{1}{2\beta}p^ip^i -\frac{1}{2}\beta z^2 -\beta) z^{E_0
- h_1 -\frac{3}{2}} \beta^{E_0 -1}|v_0\rangle\,, \ee where
$|v_0\rangle$ depends only on $\zeta$. To find a dependence of
$|v_0\rangle$ on $\zeta$ we use the 2nd equation in \rf{11n}
which with the help of \rf{spiope3} gives

\be\label{v11} |v_0\rangle = \zeta^{h_1 -h_2}\,.\ee Formulas
\rf{v10},\rf{v11} give explicit representation for the ground
state which we use to find desired constraint on $E_0$
\rf{e01},\rf{e02}. To this end we analyze Eqs. \rf{14} which for
$i=x$ and $i=\bar{x}$ lead to the equations

 \be
\label{con21}(E_0 -h_1 -1 +a +\zeta\partial_\zeta)M^{zx}
|v_0\rangle =0\,, \ee \be\label{con22} (E_0 -h_1 -1 - a
-\zeta\partial_\zeta)M^{z\bar{x}}|v_0\rangle =0\,. \ee From now on
we analyze the cases $h_1>|h_2|$ and $h_1=|h_2|$ in turn.

i) If $h_1 > |h_2|$ then Eqs. \rf{con21},\rf{con22} give

\be E_0 -2 + a -h_2 =0\,, \qquad E_0 -2h_1-2 - a + h_2 =0\,.\ee It
is easy to see that solution to these equations is $E_0=h_1+2$,
$a = h_2-h_1$. These $E_0$ and $a$ correspond to massless fields
(see \rf{e01},\rf{c12},\rf{adsmas2}).

ii) If $h_1 = h_2$, $h_2>0$ then Eq. \rf{con21} is satisfied
automatically while Eq. \rf{con22} gives $a=E_0-2 -h_1$.

iii)  If $h_1 = -h_2$, $h_2<0$ then Eq. \rf{con22} is satisfied
automatically while Eq. \rf{con21} gives $a=2-E_0 -h_1$. Thus the
cases ii) and iii) lead to values of $a$ corresponding to the
self-dual massive fields (see \rf{adsmas2},\rf{c12}). As to
restriction on $E_0$ \rf{e02} it follows simply from the
requirement that the ground state \rf{v10} be square-integrable
with respect to $z$: $\int dz ||v^{E_0,\bf{h}}\rangle|^2<
\infty$. A requirement of square-integrability with respect to
remaining variables does not lead to new constraints.

\medskip
{\bf AdS/CFT correspondence}. In spite of its Lorentz
noncovariance, a light-cone formalism offers conceptual and
technical simplification of approaches  to various problems of
modern quantum field theory. One of problems which allows to
demonstrate efficiency of a light-cone approach is $AdS/CFT$
correspondence \cite{malda,gub,wit}. In the Lorentzian signature
$AdS/CFT$ correspondence at the level of state/operator matching
between the bulk massless arbitrary spin totally symmetric fields
and the corresponding boundary operators was demonstrated in
\cite{rrm1}. Here we would like to demonstrate this
correspondence in the Euclidean signature at the level of two
point function\footnote{Discussion of $AdS/CFT$ correspondence
for spin one Maxwell field, $s=1$, and graviton, $s=2$, may be
found in \cite{wit} and \cite{liu,aru} respectively.}. In
light-cone gauge the massless arbitrary spin field in $AdS_d$ is
described by an totally symmetric tracesless $so(d-2)$ tensor
field $\phi^{I_1\ldots I_s}$, $I=1,\ldots, d-2$. This tensor can
be decomposed into traceless totally symmetric tensors of
$so(d-3)$ algebra $\phi^{I_1\ldots I_s} =\sum_{s^\prime =0}^s
\phi_{s^\prime}^{i_1\ldots i_{s^\prime}}$, $i=1,\ldots,d-3$. The
Euclidean light-cone gauge action\footnote{Note that only in this
section we use the Euclidean signature. For $d=4$ the $AdS$ mass
operator is equal to zero for all massless fields. Here we
restrict our attention to the dimensions $d\geq 5$.} takes then
the following simple form \cite{rrm1,rrm2}

\be\label{acteuc} S_{l.c.}^E =\sum_{s^\prime=0}^s \frac{1}{2}\int
d^dx( |d \phi_{s^\prime}|^2
+\frac{1}{z^2}A_{s^\prime}|\phi_{s^\prime}|^2)\,, \ee where an
eigenvalue of the $AdS$ mass operator for $\phi^{i_1\ldots
i_{s'}}$ is given by

\be A_{s^\prime} = \nu_{s'}^2 -\frac{1}{4}\,,\qquad \nu_{s'}\equiv
s^\prime +\frac{d-5}{2}\,. \ee Attractive feature of this action
is that there are no contractions of tensor indices of fields
with the ones of space derivatives, i.e., the action looks like a
sum of actions for `scalar' fields with different mass terms. This
allows us to extend the analysis of Ref.\cite{wit} in a rather
straightforward way. Using Green's function method a solution to
equations of motion

\be \Bigl(-\partial_{\bf x}^2 -\partial_z^2
+\frac{1}{z^2}(\nu_{s'}^2-\frac{1}{4})\Bigr)\phi^{i_1\ldots
i_{s'}}({\bf x},z) = 0\,,\qquad {\bf x}\equiv (x^1,\ldots,
x^{d-1})\ee is found to be

\be \phi^{i_1\ldots i_{s'}}({\bf x},z) =\int d{\bf x}'
\frac{z^{\nu_{s'}+\frac{1}{2}}}{(z^2 +|{\bf x} -{\bf
x}'|^2)^{\nu_{s'}+\frac{d-1}{2}}}\tilde{{\cal O}}^{i_1\ldots
i_{s'}}({\bf x}')\,.\ee As was expected \cite{rrm1} this solution
behaves for $z \rightarrow 0$ like
$z^{-\nu_{s'}+\frac{1}{2}}\tilde{\cal O}({\bf x})$. Plugging this
solution into the action \rf{acteuc} and evaluating a surface
integral gives

\be S_{l.c.}^E = \sum_{s'=0}^s\frac{1}{2}(\nu_{s'}+\frac{1}{2})
\int d{\bf x}d{\bf x}' \frac{\tilde{{\cal O}}^{i_1\ldots
i_{s'}}({\bf x})\tilde{{\cal O}}^{i_1\ldots i_{s'}}({\bf
x}')}{|{\bf x} -{\bf x}'|^{2\nu_{s'}+d-1}}\ee This is $so(d-3)$
decomposition of two point function of spin $s$ conserved current.

{}For the case of $AdS_5$ this analysis can be generalized to
arbitrary $D(E_0, h_1,h_2)$ fields \rf{e01},\rf{e02} in a rather
straightforward way. The most convenient way to do this is to use
representation for $AdS$ mass operator given in \rf{adsmas2}. All
that is required is to replace in above formulas the $\nu_{s'}$ by
$ \hat{\nu} = |\zeta\partial_\zeta + a|$, and the fields
$\phi_{s'}$ by $|\phi\rangle$ \rf{zetexp}. Repetition of the
procedure  outlined above gives  \be S_{l.c.}^E = \frac{1}{2}\int
d {\bf x}d {\bf x}'\ \langle\tilde{{\cal O}}({\bf x})|
\frac{\hat{\nu}+\frac{1}{2}}{|{\bf x} -{\bf
x}'|^{2\hat{\nu}+4}}|\tilde{{\cal O}}({\bf x}')\rangle\,.\ee As in
\rf{zetexp} the ket-notation $|{\cal O} \rangle$ is used to
indicate dependence of conformal operator on the complex-valued
variable $\zeta$. A scalar product $\langle {\cal O} |{\cal
O}'\rangle$ is defined as in \rf{scapro}.

\bigskip {\bf Acknowledgments}. This work was supported by the
INTAS project 991590, by the RFBR Grant No.02-02-17067, and RFBR
Grant for Leading Scientific Schools, Grant No. 01-02-30024.

\bigskip {\bf Appendix A}. Generic commutation relations of
$so(4,2)$ algebra generators $J^{CF}$, $C,F=0',0,1\ldots, 4$, we
use are

\be \label{genbas} [J^{AB},J^{CF}]=\eta^{BC}J^{AF}+ 3\hbox{
terms}\,, \quad \eta^{CF}=(-,-,+,\ldots, +)\,.\ee In the conformal
algebra notation the $J^{CF}$ split into translations $P^a$,
conformal boosts $K^a$, dilatation $D$, and $so(3,1)$ rotations
$J^{ab}$, $a,b=0,1,2,3$. Commutation relations in this basis are
\begin{eqnarray}
\label{ppkk}  [D,P^a]=-P^a\,, \qquad [D,K^a]=K^a\,, \qquad
[P^a,K^b]=\eta^{ab}D-J^{ab}\,,
\\
{} [P^a,J^{bc}]=\eta^{ab}P^c -\eta^{ac}P^b\,, \qquad
[K^a,J^{bc}]=\eta^{ab}K^c-\eta^{ac}K^b\,,
\\
\label{pkjj} [J^{ab},J^{ce}]=\eta^{bc}J^{ae}+3\hbox{ terms}\,,
\qquad \eta^{ab}=(-,+,+,+)\,.
\end{eqnarray}
In the light-cone basis these generators split into two groups:

\be P^+,\ \ P^i,\ \ K^+,\ \ K^i,\ \ J^{+i},\ \ J^{+-},\ \
J^{ij},\ \ D, \ee which we refer to as kinematical generators, and
$P^-$, $K^-$, $J^{-i}$ which we refer to as dynamical generators.
The kinematical generators have positive or zero $J^{+-}$
charges, while dynamical generators have negative $J^{+-}$
charges. Commutation relations in light-cone basis are obtainable
from \rf{ppkk}-\rf{pkjj} with the light-cone metric having the
following non-vanishing elements $\eta^{+-}=\eta^{-+}=1$,
$\eta^{ij}=\delta^{ij}$.

In energy basis the generators $J^{CF}$ split into energy operator
$J^{w\bar{w}}$, deboost operators $J^{wy}$, $J^{w\bar{y}}$,
$J^{wi}$, boost operators $J^{\bar{w}y}$, $J^{\bar{w}\bar{y}}$,
$J^{\bar{w}i}$, and $so(4)$ algebra generators $J^{iy}$,
$J^{i\bar{y}}$, $J^{y\bar{y}}$, $J^{ij}$. Commutation relations
in this basis are obtainable from \rf{genbas} with the metric
having the following non-vanishing elements
$\eta^{x\bar{x}}=\eta^{y\bar{y}}=-\eta^{w\bar{w}}=1$ (in the
$so(2)$ notation $\eta^{ij}=\delta^{ij}$). Inter-relation of the
generators in light-cone and energy bases is given by
\begin{eqnarray}
\label{in1}&& J^{w\bar{w}} = \frac{1}{2}( P^+ - P^- + K^+ -
K^-)\,,
\\
\label{in2} && J^{y\bar{y}} = \frac{1}{2}( P^+ + P^-  - K^+ -
K^-)\,,
\\
\label{in3} && J^{wy} = \frac{1}{2}( P^- + K^+  + D + J^{+-})\,,
\\
\label{in4} && J^{w\bar{y}} = \frac{1}{2}( P^+ + K^- - D  +
J^{+-})\,,
\\
\label{in5} && J^{\bar{w}y} = \frac{1}{2}( - P^+ -  K^- - D +
J^{+-})\,,
\\
\label{in6} &&  J^{\bar{w}\bar{y}} = \frac{1}{2}( -P^- - K^+  + D
+ J^{+-})\,,
\\
\label{in7} && J^{wi} = \frac{1}{2}( P^i + K^i + J^{+i} -
J^{-i})\,,
\\
\label{in8} && J^{\bar{w}i} = \frac{1}{2}( -P^i - K^i + J^{+i} -
J^{-i})\,,
\\
\label{in9} && J^{yi} = \frac{1}{2}( P^i - K^i + J^{+i} +
J^{-i})\,,
\\
\label{in10} && J^{\bar{y}i} = \frac{1}{2}( -P^i + K^i + J^{+i} +
J^{-i})\,. \end{eqnarray}

{\bf Appendix B}. In this appendix we explain a procedure of
derivation of the $AdS$ mass operator and the operator $B$ which
enter  in the definition of the light-cone gauge action and
representation of the generators of $so(4,2)$ algebra on physical
fields. As was demonstrated in \cite{rrm1} the operators $A$ and
$B$ should satisfy the basic defining equations
\begin{eqnarray}
\label{defconinv}
&& {} [A,M^{ij}]=0\,,\\
\label{defcon3} && 2\{M^{zi},A\}-[[M^{zi},A],A]=0\,,
\\
\label{defcon4} && -[M^{zi},[M^{zj},A]] + \{M^{il},M^{lj}\} +
\{M^{zi},M^{zj}\} =-2\delta^{ij}B\,.
\end{eqnarray}
To find solution of these equations we use representation for
spin operators $M^{ij}$, $M^{zi}$ in terms of projective
complex-valued variable $\zeta$ given in
\rf{spiope3},\rf{spiope4}. In general the $A$ is an operator
depending on $\zeta$ and $\partial_\zeta$. Equation
\rf{defconinv} tells that the $A$ depends only on $\hat{N}$,
where $\hat{N}\equiv \zeta\partial_\zeta$. We now are looking for
$A=A(\hat{N})$ satisfying the remaining equations
\rf{defcon3},\rf{defcon4}. Making use of the obvious relations \be
A(\hat{N})\partial_\zeta =\partial_\zeta A(\hat{N}-1)\,, \qquad
A(\hat{N})\zeta=\zeta A(\hat{N}+1)\,,\ee we get from Eqs.
\rf{defcon3} for $i=x$ and $i=\bar{x}$ the respective functional
equations
\begin{eqnarray}
&& 2(A(\hat{N})+A(\hat{N}-1))-(A(\hat{N})-A(\hat{N}-1))^2=0\,,
\\
&& 2(A(\hat{N})+A(\hat{N}+1))-(A(\hat{N})-A(\hat{N}+1))^2=0\,.
\end{eqnarray}
It is easy to see that solution of these equations does not
involve higher than second-order terms in $\hat{N}$. Making use of
general anzats $A=q\hat{N}^2 + r\hat{N}+ t$ we find that $q=1$,
$r=2a$, $t =a^2-1/4$, where $a$ is arbitrary constant. Thus we
get $A=(\hat{N}+a)^2-1/4$. To find the operator $B$ we use Eq.
\rf{defcon4} for $i=x$, $j=\bar{x}$. Solution to this equation is
found to be \be\label{bope2} B= - \{M^{zx},\,M^{z\bar{x}}\} -
(a+h_1)M^{x\bar{x}}\,. \ee The remaining unknown number $a$ can
be fixed in various ways. One way to fix $a$ is to consider the
ground state (see Section below Eq.\rf{spiope4}). Here we would
to outline procedure of finding $a$ by using technique of Casimir
operators. Let $Q$ be the second order Casimir operator. Making
use of expressions for generators \rf{3spi}-\rf{genkp} the
equation $(Q-\langle Q \rangle)|\phi\rangle=0$, where $\langle
Q\rangle$ is the eigenvalue of the $Q$ in $D(E_0,{\bf h})$ \be
-\langle Q \rangle=E_0(E_0-4)+h_1(h_1+2)+h_2^2\,,\ee can be cast
into the form (for details see Section 4 in Ref.\cite{rrm1})
\be\label{lccas} -A+2B+\frac{1}{2}M_{ij}^2+\frac{15}{4} -\langle
Q \rangle =0\,. \ee Plugging solution for $A$ and $B$ into
\rf{lccas} we obtain the equation \be\label{qeq} (E_0-2)^2 -h_1^2
+h_2^2 - (a+h_1)^2=0\,. \ee In the same way an evaluation of the
third order Casimir operator gives the equation $h_1(h_1 +a)
=(E_0-2)h_2$. For massive fields \rf{e02} this equation and
\rf{qeq} lead  to $a$ given in \rf{adsmas2},\rf{c12}. For the
case of $h_1>|h_2|$ we get $E_{0\pm}=2\pm h_1$ and the physical
relevant value $E_{0+}$ \rf{e01} can be fixed by unitarity
condition (see e.g. \cite{metsit1,brimet}). Making use of
normalization for Levi-Civita tensor in complex frame
$\epsilon^{x\bar{x}}=-{\rm i}$ it is easy to see that the
operators $A$ \rf{adsmas2} and $B$ \rf{bope2} can be cast into
$so(2)$ covariant form \rf{adsmas1},\rf{bope1}. The most
convenient way to relate the spin operators in $so(2)$ covariant
form \rf{spiope1n},\rf{spiope2n} and the ones in
\rf{spiope3n},\rf{spiope4} is to pass from the oscillators
\rf{bososc} to complex-valued variables $a^i$, $a^z$ subject to
the constraint $a^ia^i-a^za^z=0$. For simplicity we consider the
bosonic case. In terms of $a^i$, $a^z$ the $so(2)$ covariant form
of the spin operators is \be\label{spiope6}
M^{ij}=a^i\partial_{a^j}-a^j\partial_{a^i}\,,\qquad
M^{zi}=a^z\partial_{a^i} + a^i\partial_{a^z}\,. \ee Introducing
the $\zeta=a^z/a^x$ one can make sure that a representation of the
operators \rf{spiope6} on a monomial having $h_1$th power in
$a^i$, $a^z$ takes the same form as in \rf{spiope3n},\rf{spiope4}.

\end{document}